\documentstyle[12pt,epsfig]{article}

\def\la{\mathrel{\mathpalette\fun <}}

\def\fun#1#2{\lower3.6pt\vbox{\baselineskip0pt\lineskip.9pt
\ialign{$\mathsurround=0pt#1\hfil##\hfil$\crcr#2\crcr\sim\crcr}}}

\begin{document}
\baselineskip=21pt

\begin{titlepage} 

\date{ } 

\vskip 4 cm 

\title{\bf MONOJET RATES IN ULTRARELATIVISTIC HEAVY ION COLLISIONS} 

\vskip 1cm 

\author{\Large I.P.Lokhtin, L.I.Sarycheva, A.M.Snigirev \\ ~ \\ 
\it 119899, Moscow State University, Nuclear Physics Institute, Moscow, Russia}

\maketitle 

\vskip 2cm   

\begin{abstract} 
Basing on model concepts we research the monojet-to-dijet ratio as a function of the 
jet energy detection threshold in ultrarelativistic collisions of nuclei. 
We provide an comparative analysis of the contribution to monojet yield 
from gluon radiation before initial hard parton-parton scattering and from 
non-symmetric dijet energy losses in quark-gluon plasma, which expected to be created 
in ultrarelativistic heavy ion collisions. 

\vskip 7cm  

\noindent 
\underline{\hspace{8cm}} \\ 
Talk given at XIVth Conference on Ultrarelativistic Nucleus-Nucleus Collisions \\  
"Quark Matter'99", Torino, Italy, May 10-15, 1999

\end{abstract} 
\end{titlepage}

Hard jet production is considered to be an effective probe for formation of
super-dense matter -- quark- gluon plasma (QGP) in future heavy ion collider experiments
at RHIC and LHC. High $p_T$ parton pair (dijet) from a single hard scattering is 
produced at the initial stage of the collision process (typically, at $\la 0.01$ fm/c). 
It then propagates through the QGP formed due to mini-jet production at larger time 
scales ($\sim 0.1$ fm/c), and interacts strongly with the comoving constituents in the 
medium. The various aspects of hard parton passage through the dense matter are
discussed intensively~\cite{appel,blaizot,heinz,ryskin,gyul90,thoma91,gyul94,gyulqm95,baier,zakharov,lokhtin1,lokhtin2}. 
In particular, the strong  
acoplanarity of dijet transverse momentum~\cite{appel,blaizot,heinz}, the dijet 
quenching (a suppression of high $p_T$ jet pairs)~\cite{gyul90} and a monojet-to-dijet 
ratio enhancement~\cite{gyulqm95} were originally proposed as possible signals of dense 
matter formation in ultrarelativistic ion collisions. 

In the simple QCD picture for a single hard parton-parton scattering without initial
state gluon radiation (i.e. when jets from dijet pair escape from primary hard 
scattering vertex back-to-back in azimuthal plane with equal absolute transverse 
momentum values, $p_{T1} = p_{T2}$) a monojet is created only if one of the two hard 
partonic jets loses so much energy due to multiple scattering in the dense matter that 
effectively we can detect only one single jet in the final state. The monojet 
rate is obtained by integrating the dijet rate over the transverse
momentum $p_{T2}$ of the second (unobserved) jet with the condition that $p_{T2}$ 
be smaller than the threshold value $p_{cut}$ (or the threshold jet energy $E_T =
p_{cut}$~\footnote{Due to fluctuations of the transverse energy flux arising from a
huge multiplicity of secondary particles in the event, the "true" jet recognition in 
ultrarelativistic heavy ion collisions is
possible beginning only from some energy threshold~\cite{kruglov}.}). Then rate of 
dijets $R^{dijet}$ with $p_{T1}, p_{T2} > p_{cut}$ and monojets $R^{mono}$ with 
$p_{T1} > p_{cut}$ ($p_{T2} < p_{cut}$) in central $AA$ collisions is calculated as 
integral over all possible jet transverse momenta $p_{T1}$, $p_{T2}$ and longitudinal 
rapidites $y_1$, $y_2$. 

At first in the framework of the simple model~\cite{lokhtin3} we demonstrate that
monojet-to-dijet ratio can be related to mean the acoplanarity measured in the units 
of the jet threshold energy, namely 
\begin{equation}
\frac{R^{mono}}{R^{dijet}} \propto \frac{<|K_T|>}{E_T}.   
\end{equation}

\begin{figure}[hbtp] 
\begin{center} 
\makebox{\epsfig{file=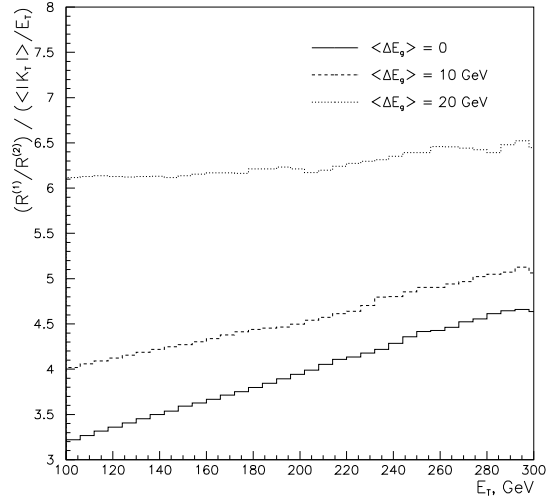, height=80mm}}   
\caption{(Monojet/dijet)/(mean acoplanarity/$E_T$) as function of jet energy threshold 
$E_T$.}
\end{center} 
\end{figure}

The results of physics simulation have been obtained in the three scenarios for jet
quenching due to collisional energy losses~\footnote{Although the radiative energy 
losses of a high energy parton dominate over the collisional losses by up to an order 
of magnitude, it will, in the first place, soften particle energy distributions inside 
the jet, increase the multiplicity of secondary particles, but will not affect the 
total jet energy~\cite{lokhtin2}. On the other hand, the collisional energy loss turns 
out to be practically independent on jet cone size and emerges outside the narrow jet 
cone.} of jet partons in mid-rapidity region 
$y=0$~\cite{lokhtin1,lokhtin3}: $(i)$ no jet quenching, $(ii)$ jet quenching in a 
perfect longitudinally expanding QGP (the average collisional energy losses of a hard 
gluon $<\Delta E_{g}> \simeq 10$ GeV, $<\Delta E_{q}> = 4/9 \cdot <\Delta E_{g}>$), 
$(iii)$ jet quenching in a maximally viscous quark-gluon fluid, resulting in 
$<\Delta E_{g}> \simeq 20$ GeV. Initial state gluon radiation has been taken into 
account with the PYTHIA Monte-Carlo model~\cite{pythia} at c.m.s. energy 
$\sqrt{s} = 5.5 A$ TeV. 

Then we conclude that rescattering of hard partons in medium results in weaker
$E_T$-dependence of ratio $R^{mono}/R^{dijet}$ to $<|K_T|>/E_T$ (see fig.1). With 
growth of energy
losses the ratio we are interested in has a tendency to be constant, what would be
interpreted as the signal of super-dense matter formation.


\newpage

\begin{thebibliography}{99}
\bibitem{appel} D.A.Appel, Phys. Rev. D33 (1986) 717 
\bibitem{blaizot} J.P.Blaizot, L.D.McLerran, Phys. Rev. D34 (1986) 2739 
\bibitem{heinz} M.R.Rammerstorfer, U.Heinz, Phys. Rev. D41 (1990) 306 
\bibitem{ryskin} M.G.Ryskin, Sov. J. Nucl. Phys. 52 (1990) 139 
\bibitem{gyul90} M.Gyulassy, M.Plumer, Phys. Lett. B 234 (1990) 432
\bibitem{thoma91} M.H.Thoma, Phys. Lett. B 273 (1991) 128
\bibitem{gyul94} M.Gyulassy, X.-N.Wang, Nucl. Phys. B420 (1994) 583; 
X.-N.Wang, M.Gyulassy, M.Plumer, Phys. Rev. D51 (1995) 3436   
\bibitem{gyulqm95} M.Plumer, M.Gyulassy, X.-N.,Wang, Nucl. Phys. A590 (1995) 511  
\bibitem{baier} R.Baier, Yu. L.Dokshitzer, S.Peigne, D.Schiff, Phys. Lett B 345 
(1995) 277; R.Baier, Yu. L.Dokshitzer, A.H.Mueller, S.Peigne, D.Schiff,   
Nucl. Phys. B 483 (1997) 291; 484 (1997) 265; R.Baier, Yu. L.Dokshitzer, 
A.H.Mueller,  D.Schiff,  Nucl. Phys. B 531 (1998) 403; Phys. Rev. C58 (1998) 1706   
\bibitem{zakharov} B.G.Zakharov, JETP Lett. 65 (1997) 615      
\bibitem{lokhtin1} I.P.Lokhtin, A.M.Snigirev, Phys. At. Nucl. 60 (1997) 360; 
Z.Phys C73 (1997) 315 
\bibitem{lokhtin2} I.P.Lokhtin, A.M.Snigirev, Phys. Lett. B 440 (1998) 163   
\bibitem{kruglov} N.A.Kruglov, I.P.Lokhtin, L.I.Sarycheva, A.M.Snigirev, Z. Phys. C 
76 (1997) 99 
\bibitem{lokhtin3} I.P.Lokhtin, L.I.Sarycheva, A.M.Snigirev, Phys. Atom. Nucl. 62 
(1999), in press 
\bibitem{pythia} T.Sjostrand, Comp. Phys. Com. 82 (1994) 74  
\end{thebibliography}
\end{document}